\documentclass[onecolumn]{revtex4}
\usepackage{color}
\usepackage{graphicx}
\usepackage{dcolumn}
\usepackage{color}
\usepackage{bm}
\usepackage{amsmath}
\usepackage{amsfonts}
\usepackage{amssymb}
\RequirePackage[colorlinks,citecolor=red,urlcolor=magenta,linkcolor=blue]{hyperref}
\def\eq#1{{Eq.~(\ref{#1})}}
\def\eqs#1{{Eqs.~(\ref{#1})}}
\def\EH{general relativity }
\def\LL{Lanczos-Lovelock }
\def\EHA{Einstein-Hilbert action}
\input epsf

\begin{document}

\title{Evolution of Spacetime arises due to the departure from Holographic Equipartition in all \LL Theories of Gravity}

\author{Sumanta
Chakraborty \footnote{sumanta@iucaa.ernet.in;\\
sumantac.physics@gmail.com}}

\affiliation{IUCAA, Post Bag 4, Ganeshkhind,
Pune University Campus, Pune 411 007, India}

\author{T. Padmanabhan \footnote{paddy@iucaa.ernet.in}}

\affiliation{IUCAA, Post Bag 4, Ganeshkhind,
Pune University Campus, Pune 411 007, India}

\date{\today}

\begin{abstract}

In the case of general relativity one can interpret the Noether 
charge in any bulk region as the heat content $TS$ of its  boundary surface. Further,  the time evolution of spacetime metric in Einstein's theory arises due to the difference $(N_{sur}-N_{bulk})$ of suitably defined surface and bulk degrees of freedom. We show that this thermodynamic interpretation generalizes in a natural fashion to all \LL\ models of gravity. The Noether charge, related to time evolution vector field, in a bulk region of space is equal to the heat content $TS$ of the boundary surface with the temperature $T$ defined using local Rindler observers and $S$ being the Wald entropy. Using the Wald entropy to define the surface degrees of freedom $N_{sur}$ and Komar energy density to define the bulk degrees of freedom $N_{bulk}$, we can also show that the time evolution of the geometry is sourced by $(N_{sur}-N_{bulk})$.
When it is possible to choose the foliation of spacetime such that metric is independent of time, the above dynamical equation 
yields the holographic equipartition for \LL gravity with $N_{sur}=N_{bulk}$. The implications are discussed.

\end{abstract}

\maketitle

\section{Introduction}\label{Paper1:Sec:Intro}

A surprising connection between gravity and thermodynamics 
 was first demonstrated in the context of black hole mechanics by the fact that one can associate an entropy  \cite{Bekenstein1973,Bekenstein1974} and temperature \cite{Hawking1975,Hawking1976}, with the black holes. It was soon realised that similar connection exists in the case of several other horizons \cite{Davies1976,Unruh1976} and that the ideas have a far greater domain of applicability \cite{Padmanabhan2005a,Wald2001}.

Further work in the last decade suggests that these results could be just the tip of the iceberg. It appears that (i) gravitational field equations themselves may have only the status similar to equations in other emergent phenomenon like in kinetic theory of gases or fluid mechanics \cite{Padmanabhan2010a,Sakharov1968,Jacobson1995,Volovik2001} and (ii) this emergent interpretation  is applicable to 
theories more general than Einstein gravity and has a universal nature.
Some of the results which lend support to this  paradigm are the following:
\begin{itemize}
 \item 
 The gravitational field equations reduce to thermodynamic identities 
on horizons for a wide class of gravity theories more general than Einstein gravity 
\cite{Padmanabhan2002,Cai2005,Padmanabhan2006a,Akbar2006,Padmanabhan2006b,Padmanabhan2009}. 
\item
The action describing gravity can be separated into a bulk term and 
a surface term with a specific (`holographic') relation between them,
not only in Einstein gravity but also in more general class of theories 
\cite{Padmanabhan2010b,Padmanabhan2006c,Kolekar2010,Kolekar2012b}. In fact the action functional in all \LL gravity can be given a thermodynamic interpretation \cite{Kolekar2012b,Padmanabhan2005b,Padmanabhan2010e,Hawking1977}.
\item
Gravitational field equations in all \LL\ models can be obtained from thermodynamic 
extremum principles \cite{Padmanabhan2007,Padmanabhan2008} 
involving the heat density of null surfaces in the spacetime.
\item
Gravitational field equations reduce to Navier-Stokes equations of fluid 
dynamics in arbitrary spacetime projected on a null surface generalizing previous results 
on black hole spacetime \cite{Padmanabhan2011a,Kolekar2012a,Damour1982}.
\end{itemize}
More recently \cite{Padmanabhan2013a} these ideas have been taken significantly further in the context of general relativity. One of us (TP) demonstrated that, in the context of general relativity, the following results hold:
(a) The total Noether charge in a 3-volume $\mathcal{R}$, related to the time evolution vector field, can be interpreted as the heat content of the boundary $\partial \mathcal{R}$ of the volume. This provides yet another holographic result connecting the bulk and boundary variables.
(b) The time evolution of the spacetime itself can be described in an elegant manner by the equation:
 \begin{equation}
 \int_\mathcal{R}\frac{d^3x}{8\pi}h_{ab}\pounds_\xi p^{ab}  
 = \epsilon\frac{1}{2} k_B T_{\rm avg} ( N_{\rm bulk} - N_{\rm sur})
\label{start}
\end{equation} 
where $h_{ab}$ is the induced metric on the $t=$ constant surface, the $p^{ab}$ is its conjugate momentum, $\xi^a=Nu^a$ is the time evolution vector corresponding to observers  with four-velocity $u_a  = - N \nabla_a t$ that is the normal to the $t=$ constant surface. The $N_{\rm sur}$ and $N_{\rm bulk}$ are the degrees of freedom in the surface and bulk of a 3-dimensional region $\mathcal{R}$ and $T_{\rm avg}$ is the average Davies-Unruh temperature   of the boundary. (The parameter $\epsilon=\pm 1$ ensures that the $N_{bulk}$ is positive even when Komar energy turns negative.) This equation shows that the  rate of change of gravitational momentum  is driven by the departure from holographic equipartition, measured by  $(N_{\rm bulk} - N_{\rm sur})$. The metric will be time independent in the chosen foliation if  $N_{\rm sur} = N_{\rm bulk}$ which can happen for all static geometries.  The validity of \eq{start} for all observers (i.e., foliations) implies the validity of Einstein's equations. In short, \textit{deviation from holographic equipartition leads to the time evolution of the metric}.

In the past, virtually every result indicating the emergent nature of gravity in the context of general relativity could be generalized to all \LL models of gravity. It is therefore worth investigating whether the above description can be generalized to \LL models. This is very important because the expression for horizon entropy in general relativity is rather trivial and is just a quarter of horizon area. In \LL models, the corresponding expression is much more complex which, in turn, modifies the expression for $N_{sur}$. It is, therefore, not clear a priori whether our results --- interpretation of Noether charge and \eq{start} ---  will generalize to \LL models. 
We will show here that, these results indeed
possess a natural generalization to \LL gravity as well.

The rest of the paper is organised as follows: In Sec. \ref{Paper1:Sec:EHEqui} we review the know results for Einstein gravity and clarify some technical points. (In particular, in Sec \ref{Paper1:Sec:EHEqui:Application} we give some explicit examples to illustrate what happens when the same spacetime admits both  static and nonstatic folications.) In Sec. \ref{Paper1:Sec:LLEqui} we generalise all these results to \LL models of gravity. Sec. \ref{Paper1:Sec:LLEqui:Intro} provides a brief introduction to \LL models and set up the notation etc. In Sec \ref{Paper1:Sec:LLEqui:Heat} we relate the Noether charge to the surface heat content in the \LL models and in Sec \ref{Paper1:Sec:LLEqui:Evol} we derive the evolution equation in terms of surface and bulk degrees of freedom. The last section summarises the conclusions. We work with a mostly positive signature in D dimensional spacetime and use units with $G=\hbar=c=1$.
\section{Warm up: Review of the results for Einstein gravity}
\label{Paper1:Sec:EHEqui}

\subsection{The foliation of spacetime}\label{Paper1:Sec:EHEqui:Setup}

We start with a spacetime foliated by a series of spacelike hypersurfaces 
each being determined by the constant value of a scalar field $t(x)$, such that 
on each hypersurface $t(x)=\textrm{constant}$. The unit normal to the constant $t(x)$
hypersurface is $u_{a}=-N\nabla _{a}t$, which reduces to $-N\delta ^{0}_{a}$ 
when $t$ is considered as one of the coordinates in this spacetime. For this spacetime 
foliation we have $g^{00}=-1/N^{2}$, and $u^{a}u_{a}=-1$. 
Given such a foliation,  we can introduce a time evolution vector 
$\zeta ^{a}$ by the condition $\zeta ^{a}\nabla _{a}t=1$, which 
in the coordinate system with $t$ as a coordinate becomes $\zeta ^{a}=\delta ^{a}_{0}$. 
In general, we can readily obtain the following decomposition: 
$\zeta ^{a}=-\left(\zeta ^{b}u_{b}\right)u^{a}+N^{a}$, with the property 
$N^{a}u_{a}=0$ and $N^{a}=h^{a}_{b}\zeta ^{b}$, where $h^{a}_{b}=\delta ^{a}_{b}+u^{a}u_{b}$ 
being the projection tensor. 
This decomposition also introduces another vector 
\begin{equation}\label{Paper1:Eq01}
\xi _{a}=Nu_{a}\to-N^{2}\delta ^{0}_{a}
\end{equation}
where the last result holds in the preferred foliation.
If we impose 
the coordinate condition that $t$ becomes one of the spacetime coordinate 
and $g_{0\alpha}=0$ this vector 
reduces to $\zeta ^{a}$. Further, in static spacetimes $\xi ^{a}$ turns out to be the time-like 
Killing vector. It was shown in ref. \cite{Padmanabhan2013a} that this vector plays a crucial role in the thermodynamic interpretation and
has a rich structure as far as the Noether current and spacetime 
dynamics is concerned. 
\subsection{Noether Charge and evolution equation in \EH\ }
\label{Paper1:Sec:EHEqui:Evolution}

We begin by calculating the Noether charge  for 
the vector field $\xi ^{a}$. The Noether current in \EH can be written in 
an elegant manner using a new set of variables $(f^{ab}, N^{c}_{ab})$ in terms of which several expressions in 
general relativity becomes simpler. These variables, defined as: 
\begin{equation}
f^{ab}=\sqrt{-g}g^{ab};\qquad N^{c}_{ab}=Q^{cd}_{ae}\Gamma ^{e}_{bd}+Q^{cd}_{be}\Gamma ^{e}_{ad}
\end{equation} 
where  $2Q^{ab}_{cd}=(\delta ^{a}_{c}\delta ^{b}_{d}-\delta ^{a}_{d}\delta ^{b}_{c})$ were earlier used in 
\cite{Babak1999,Kijowski1997} and their thermodynamic interpretation was provided in \cite{Krishna2013}. The variation of the \EHA\ 
in terms of these conjugate variables results into:
\begin{eqnarray}\label{Paper1:Eq02}
\delta \left(\sqrt{-g}R\right)&=&R_{ab}\delta f^{ab}-\partial _{c}\left(f^{ab}\delta N^{c}_{ab}\right)
\nonumber
\\
&=&\sqrt{-g}\left[G_{ab}\delta g^{ab}-\nabla _{c}\left(g^{ik}\delta N^{c}_{ik}\right)\right]
\end{eqnarray} 
If the above variation results from a Lie variation with respect to some 
vector field $v_{a}$ then from the above expression a conserved current $J^{a}$ emerges 
with the property $\nabla _{a}J^{a}=0$. This conserved current is the Noether current and has 
the following expression:
\begin{equation}\label{Paper1:Eq03}
16\pi J^{a}(v)=2R^{ab}v_{b}+g^{ij}\pounds _{v}N^{a}_{ij}
\end{equation}
(The factor $16\pi$ is conventional when we use units with $G=1$; obviously any multiple of $J^a$ is conserved.) Given the fact that $\nabla _{a}J^{a}=0$, we can write the Noether 
current in terms of an antisymmetric second rank tensor $J^{ab}$, the Noether potential as $J^a=\nabla_b J^{ab}$. 
This, in the case of \EH becomes:
\begin{equation}\label{Paper1:Eq04}
16 \pi J^{ab}(v)=\nabla ^{a}v^{b}-\nabla ^{b}v^{a}
\end{equation}
Though in the above discussion the Noether current has been derived using Lie variation 
it should be stressed that the same result can be obtained using differential geometry 
\emph{without ever using diffeomorphism invariance of the action principle for gravity}. 
This has been shown explicitly in ref. \cite{Padmanabhan2013a} and hence we will 
not repeat the arguments here. 

Next we will calculate the Noether current for the time evolution vector $\xi ^{a}$. 
For the evaluation we shall use a relation between Noether current of two vector fields 
$q^{a}$ and $v^{a}$ such that $v^{a}=f(x)q^{a}$, for arbitrary function $f(x)$. In
App. \ref{Paper1:App:NoetherLL} [see \eq{Paper1:EqA02}] it is shown that:
\begin{equation}\label{Paper1:Eq05}
16\pi \left\lbrace q_{a}J^{a}\left(v\right)-f(x)q_{a}J^{a}\left(q\right)\right\rbrace =
\nabla _{b}\left(\left\lbrace q^{a}q^{b}-g^{ab}q^{2} \right\rbrace \nabla _{a}f\right)
\end{equation}
The usefulness of this relation can be realized by noting that for $q_{a}=\nabla _{a}\phi$ for some 
scalar $\phi$ the Noether current vanishes. Thus applying the above result for $u_{a}$ and 
then for $\xi _{a}$ one can arrive at the following simple relation for Noether 
current of $\xi _{a}$ as 
[see App. \ref{Paper1:App:NoetherLL}; \eq{Paper1:EqA08}]:
\begin{equation}\label{Paper1:Eq06}
16 \pi u_{a}J^{a}\left(\xi \right)=2D_{\alpha}\left(Na^{\alpha}\right)
\end{equation}
where $a^{i}=u^{j}\nabla _{j}u^{i}$ represents the four acceleration 
which satisfies the relation $D_{i}a^{i}=\nabla _{i}a^{i}-a^{2}$, with $D_{i}$ 
representing the surface covariant derivative for the $t=\textrm{constant}$ surface. 
Then we can integrate \eq{Paper1:Eq06} over the $t=\textrm{constant}$ hypersurface 
with $\sqrt{h}d^{3}x$ being the integration measure and bounded 
by $N=\textrm{constant}$ surface
leading to the total Noether charge contained in the three volume. 
Then dividing both sides of \eq{Paper1:Eq06} by $16\pi$ we arrive at:
\begin{equation}\label{Paper1:Eq07}
\int _{\mathcal{V}} d^{3}x\sqrt{h}u_{a}J^{a}\left(\xi \right)
=\int _{\mathcal{V}} \frac{d^{3}x\sqrt{h}}{8\pi}D_{\alpha}\left(Na^{\alpha}\right)
=\int _{\partial \mathcal{V}}\frac{\sqrt{\sigma} d^{2}x}{8\pi}Nr_{\alpha}a^{\alpha}
\end{equation}
which holds for any arbitrary region $\mathcal{V}$ of the spacetime, with 
the bounding region being $N\left(t,\textbf{x}\right)=\textrm{constant}$ surface within 
$t=\textrm{constant}$ hypersurface. This allows us to identify the vector 
$r_{a}$ to be normal to this $N\left(t,\textbf{x}\right)=\textrm{constant}$ 
hypersurface as: 
$
r_{a}=\epsilon D _{a}N\left(D _{b}N D ^{b}N\right)^{-1/2}=
\epsilon h^{i}_{a}\nabla _{i}N/a,                                   
$
where the 
$\epsilon$ factor is introduced to ensure that $r_{a}$ is always 
the \textit{outward} pointing normal. (When the acceleration $a_i$ is outward pointing $\epsilon=1$; otherwise $\epsilon=-1$). 
Here $a=\sqrt{a_ia^i}$ is the magnitude of the acceleration. 
So we can also write the normal $r_{\alpha}$ 
as: $r_{\alpha}=\epsilon a_{\alpha}/a$, with 
$a$ representing magnitude of the acceleration. Then we obtain
\begin{equation}\label{Paper1:Eq08}
Nr_{\alpha}a^{\alpha}=N\epsilon \frac{a_{\alpha}}{a}a^{\alpha}=\epsilon Na.
\end{equation}
 The Tolman 
redshifted Davies-Unruh temperature on the boundary surface   $N=\textrm{constant}$, is $T_{loc}=Na/2\pi$ for 
observers with four velocity $u_{a}=-N\delta ^{0}_{a}$. Locally free falling observers 
will observe these observers moving normal to the $t=\textrm{constant}$ hypersurface 
with an acceleration $a$ and as a consequence the local vacuum will appear as a thermal 
state with temperature $T_{loc}$ to these observers. 
Using all these results \eq{Paper1:Eq07} can be written as:
\begin{equation}\label{Paper1:Eq09}
2\int _{\mathcal{V}} d^{3}x\sqrt{h}u_{a}J^{a}\left(\xi \right)=
\epsilon \int _{\partial \mathcal{V}}\frac{\sqrt{\sigma} d^{2}x}{2}\left(\frac{Na}{2\pi}\right)
=\epsilon \int _{\partial \mathcal{V}}\sqrt{\sigma} d^{2}x\left(\frac{1}{2}T_{loc}\right)
\end{equation}
The above result can be interpreted as: \emph{twice the Noether charge contained in the 
$N=\textrm{constant}$ surface is equal to the equipartition energy of the surface}. 
With the interpretation of $\sqrt{\sigma}/4$ as entropy density the above result also gives:
\begin{equation}\label{Paper1:Eq10}
\int _{\mathcal{V}} d^{3}x\sqrt{h}u_{a}J^{a}\left(\xi \right)=
\epsilon \int _{\partial \mathcal{V}}\frac{\sqrt{\sigma} d^{2}x}{4}T_{loc}
=\epsilon \int _{\partial \mathcal{V}}d^{2}xT_{loc}s
\end{equation}
which is the heat density of the bounding surface. The interpretation 
of $\sqrt{\sigma}/4$ as the entropy density comes naturally when the 
boundary surface becomes a horizon.  Thus, even in the most general (nonstatic) context, the  Noether charge of time development vector in the 
bulk spacetime region has a simple interpretation as 
the surface heat content. 

We will next obtain the dynamics of gravity in terms of bulk and surface degrees of freedom using 
the Noether current formalism. For this,  we again start with \eq{Paper1:Eq06} 
and use \eq{Paper1:Eq03} leading to:
\begin{equation}\label{Paper1:Eq11}
u_{a}g^{ij}\pounds _{\xi}N^{a}_{ij}=D_{\alpha}\left(2Na^{\alpha}\right)
-2NR_{ab}u^{a}u^{b}
\end{equation} 
Then we integrate the above expression as in the earlier situation over the three dimensional 
region $\mathcal{R}$ with boundary surface being $N=\textrm{constant}$ 
within the $t=\textrm{constant}$ surface leading to:
\begin{equation}\label{Paper1:Eq12}
\int _{\mathcal{R}}d^{3}x\sqrt{h}u_{a}g^{ij}\pounds _{\xi}N^{a}_{ij}
=\int _{\partial \mathcal{R}}d^{2}x\sqrt{\sigma}r_{\alpha}\left(2Na^{\alpha}\right)
-\int _{\mathcal{R}} d^{3}x\sqrt{h}2NR_{ab}u^{a}u^{b}
\end{equation}
where we have used $d^{3}x\sqrt{h}$ as the integration measure. 
Introducing the dynamics through Einstein's equation 
$R_{ab}=8\pi \left(T_{ab}-(1/2)g_{ab}T\right)=8\pi \bar{T}_{ab}$ and dividing the 
whole expression  by 
$8\pi$ gives:
\begin{equation}\label{Paper1:Eq13}
\int _{\mathcal{R}}\frac{d^{3}x\sqrt{h}}{8\pi}u_{a}g^{ij}\pounds _{\xi}N^{a}_{ij}
=\int _{\partial \mathcal{R}}d^{2}x\sqrt{\sigma}r_{\alpha}\left(\frac{Na^{\alpha}}{4\pi}\right)
-\int _{\mathcal{R}} d^{3}x\sqrt{h}2N\bar{T}_{ab}u^{a}u^{b}
\end{equation}
Using \eq{Paper1:Eq08} and introducing the Komar energy density by the definition
$\rho _{Komar}=2N\bar{T}_{ab}u^{a}u^{b}$ we obtain:
\begin{equation}\label{Paper1:Eq14}
\frac{1}{8\pi}\int _{\mathcal{R}}d^{3}x\sqrt{h}u_{a}g^{ij}\pounds _{\xi}N^{a}_{ij}=
\epsilon \int _{\partial \mathcal{R}}d^{2}x\sqrt{\sigma}\left(\frac{1}{2}T_{loc}\right)
-\int _{\mathcal{R}}d^{3}x\sqrt{h}\rho _{Komar}
\end{equation}
We  define the surface 
degrees of freedom by:
\begin{equation}\label{Paper1:Eq15}
N_{sur}\equiv A=\int _{\partial \mathcal{R}} \sqrt{\sigma}d^{2}x
\end{equation}
which is always positive. We can define an average temperature 
over the surface such that
\begin{equation}\label{Paper1:Eq16}
T_{avg}\equiv \frac{1}{A}\int _{\partial \mathcal{R}}\sqrt{\sigma}d^{2}xT_{loc}.
\end{equation}
Finally we introduce the bulk degrees of freedom by the definition:
\begin{equation}\label{Paper1:Eq17}
N_{bulk}=\frac{\epsilon}{(1/2)T_{avg}}\int d^{3}x\sqrt{h}\rho _{Komar}.
\end{equation}
When the bulk region is in equipartition  at 
the temperature $T_{avg}$ then $N_{bulk}$ represents the correct number of bulk 
degrees of freedom. Here also we need the factor $\epsilon$ to ensure that $N_{bulk}$ is positive definite. We choose $\epsilon=+1$ if the total Komar energy within the volume is 
positive and $\epsilon=-1$ if the total Komar energy in the volume is negative so as to 
keep $N_{bulk}$ always positive. 
With all these definitions \eq{Paper1:Eq14} can be written in the following manner: (This corrects a minor typo in ref. \cite{Padmanabhan2013a}.)
\begin{equation}\label{Paper1:Eq18}
\frac{1}{8\pi}\int _{\mathcal{R}}d^{3}x\sqrt{h}u_{a}g^{ij}\pounds _{\xi}N^{a}_{ij}=
\frac{\epsilon}{2}T_{avg}\left(N_{sur}-N_{bulk}\right)
\end{equation}
 Thus for comoving observers in static spacetime we have the holographic equipartition 
$
N_{sur}=N_{bulk}
$
When the difference $\left(N_{sur}-N_{bulk}\right)$ is nonzero for a given foliation,
we have departure from holographic equipartition and this 
leads to the time evolution 
of the metric, as is evident from the left hand side of \eq{Paper1:Eq18}. The implications of this result has been discussed extensively in ref. \cite{Padmanabhan2013a}.
\subsection{Aside: Some illustrative examples}
\label{Paper1:Sec:EHEqui:Application}

An important aspect of the dynamical evolution equation is the following: The structure of \eq{Paper1:Eq18} 
shows that, while it is covariant, it is foliation dependent 
through the normal $u_{i}$. For example, even in a static spacetime (which possesses a timelike Killing vector field) the \textit{non-static} observers 
will perceive a time-dependence of the metric and hence departure from holographic equipartition (so that both sides of \eq{Paper1:Eq18} are nozero), while static observers (with velocities along the Killing direction)  will perceive a time-independent metric and holographic equipartition, 
(with both sides of \eq{Paper1:Eq18} being zero).
This contrast is most striking when we study two natural class of observers in a static spacetime. The first set are observers with four-velocities along the timelike Killing vector who have a nonzero acceleration. In this foliation the metric components are independent of time and the left hand side of \eq{Paper1:Eq18} vanishes leading to 
holographic equipartition $N_{sur}=N_{bulk}$. But we know that \textit{any} spacetime metric can be expressed in the synchronous frame coordinates 
with the line element:
\begin{equation}\label{Paper1:Eq22}
ds^{2}=-d\tau ^{2}+g_{\alpha \beta}dx^{\alpha}dx^{\beta}
\end{equation}
In the synchronous frame the observers at $x^{\alpha}=\textrm{constant}$ are
comoving with four velocity:
$
u_{a}=\left(-1,0,0,0\right).
$
Obviously, the comoving observer is not accelerating, (i.e, the curves $x^\alpha=$ constant are geodesics)
and hence the local Davies-Unruh temperature for these observers will vanish. 
We want to consider \eq{Paper1:Eq18} in two such coordinate systems to clarify some of the issues.  

Let us begin with the synchronous frame in which $T_{avg}\to 0, T_{avg}N_{sur}\to0$ with 
$T_{avg}N_{bulk}$ remaining finite, so that
\eq{Paper1:Eq18} reduces to the following form:
\begin{equation}\label{Paper1:Eq21}
\frac{1}{8\pi}\int _{\mathcal{R}}d^{3}x\sqrt{h}u_{a}g^{ij}\pounds _{\xi}N^{a}_{ij}=
-\frac{\epsilon}{2}T_{avg}N_{bulk}
=-\int _{\mathcal{R}}d^{3}x\sqrt{h}\rho _{Komar}
\end{equation}
The quantity
$u_{a}g^{ij}\pounds _{\xi} N^{a}_{ij}$ in an arbitrary synchronous 
frame is given by:
\begin{eqnarray}\label{Paper1:Eq23}
\sqrt{h}u_{a}g^{ij}\pounds _{\xi}N^{a}_{ij}&=&2\sqrt{h}\left(K_{ab}K^{ab}-u^{a}\nabla _{a}K\right)
\nonumber
\\
&=&\sqrt{h}\left(g^{\alpha \beta}\partial _{\tau}^{2}g_{\alpha \beta}
+\frac{1}{2}\partial _{\tau}g^{\alpha \beta}\partial _{\tau}g_{\alpha \beta}\right)
\end{eqnarray}
where we have used \eq{Paper1:EqA12}.
It can be shown that equating this expression to $-16\pi \bar{T}_{ab}u^{a}u^{b}$ correctly reproduces the standard time-time component of Einstein's equation in the synchronous frame.
So, our \eq{Paper1:Eq18} gives the correct result, as it should.

As an explicit example, consider the Friedmann universe for which 
$g_{\alpha \beta}=a^{2}(t)\delta _{\alpha \beta}$ leading to the following expressions:
\begin{eqnarray}\label{Paper1:Eq24a}
\partial _{\tau}g_{\alpha \beta}=2a\dot{a}\delta _{\alpha \beta};~~~~
\partial _{\tau}^{2}g_{\alpha \beta}=\left(2\dot{a}^{2}+2a\ddot{a}\right)\delta _{\alpha \beta};~~~~
\partial _{\tau}g^{\alpha \beta}=-2\frac{\dot{a}}{a^{3}}\delta ^{\alpha \beta}
\end{eqnarray}
and
$\bar{T}_{ab}u^{a}u^{b}=(1/2)\left(\rho +3p\right)$. On substitution of \eq{Paper1:Eq24a}, 
in \eq{Paper1:Eq23} we 
arrive at the following expression for the time evolution of the scale factor:
\begin{equation}\label{Paper1:Eq25}
\frac{\ddot{a}}{a}=-\frac{4\pi}{3}\left(\rho +3p\right).
\end{equation}
The above equation supplemented by the equation of state  
leads to the standard results. Thus 
in Friedmann universe the dynamical evolution of spacetime leads 
to dynamical evolution equation of the scale factor sourced by the 
Komar energy density. Before proceeding further it is worthwhile to clarify the following point: In the case of Friedmann universe, one can \textit{also} obtain \cite{frwevol} the following result
\begin{equation}
 \frac{dV}{dt}=N_{sur}-\sum\epsilon N_{bulk}
\label{hvol}
\end{equation}
where $V=(4\pi/3)H^{-3}$ is the areal volume of the Hubble radius sphere if we define the degrees of freedom using the temperature $T\equiv H/2\pi$. (The $\epsilon$ factor has to chosen for each bulk component appropriately in order to keep all $N_{bulk}$ positive as indicated by the summation; see \cite{frwevol} for a detailed discussion).  Though this is also equivalent to Einstein's equation, it is structurally quite different from the evolution equation in \eq{Paper1:Eq18} (and should not be confused with it) for the following reasons: (a) The left hand sides of \eq{Paper1:Eq18} and \eq{hvol} are different. (b) The placement of $\epsilon$-s are different in the right hand sides of \eq{Paper1:Eq18} and \eq{hvol}. (c) One uses Friedmann time coordinate in the left hand side of \eq{hvol} but still attributes a temperature $T\equiv H/2\pi$ to define the degrees of freedom. (d) Most importantly, \eq{hvol} holds only for Friedmann universe while  \eq{Paper1:Eq18} is completely general.

Coming back to the consequences of \eq{Paper1:Eq18}, since this result is true for any Friedmann universe, it is also true for the de Sitter spacetime written in synchronous (Friedmann) coordinates. The de Sitter metric, as seen by comoving observers has an explicit time dependence $a(t)\propto \exp \left(Ht\right)$ and for these observers the perceived Davies-Unruh temperature vanishes. Nevertheless, \eq{Paper1:Eq18} will of course give the correct evolution equation.
 On the other hand, de Sitter spacetime can also be expressed in static coordinates with the line element:
\begin{equation}\label{Paper1:Eq26}
ds^{2}=-\left(1-\frac{r^{2}}{l^{2}}\right)dt^{2}+\frac{dr^{2}}{\left(1-\frac{r^{2}}{l^{2}}\right)}
+r^{2}\left(d\theta ^{2}+\sin ^{2}\theta d\phi ^{2} \right)
\end{equation}
The observers with $x^{\alpha}=\textrm{constant}$ in this coordinate system are not geodesic observers. They have 
the following four velocity and four acceleration respectively:
\begin{eqnarray}
u_{a}&=&\sqrt{\left(1-\frac{r^{2}}{l^{2}}\right)}\left(-1,0,0,0\right)
\label{Paper1:Eq27a}
\\
a^{i}&=&\left(0,-(r/l^{2}),0,0\right)
\label{Paper1:Eq27b}
\end{eqnarray}
Let us see what happens when we use this foliation.

In this case, the acceleration $a^i$  and the normal $r_{i}$ 
are directed opposite to each other as $r_{i}$ is the outward directed normal. 
(Note that in the de Sitter spacetime the 
free-falling observers are moving outwards and \textit{with respect to them} the static observers 
are moving \textit{inwards} opposite to the outward pointing normal.)
Hence in this situation we have $\epsilon=-1$. 
The magnitude of the acceleration is:
\begin{equation}\label{Paper1:Eq28}
a=\frac{r}{l^{2}}\frac{1}{\sqrt{\left(1-\frac{r^{2}}{l^{2}}\right)}}
\end{equation}
which is obtained from \eq{Paper1:Eq27b}. 
Thus the local Davies-Unruh temperature turns out to be:
\begin{equation}\label{Paper1:Eq29}
T_{loc}=\frac{Na}{2\pi}=\frac{r}{2\pi l^{2}}=T_{avg}
\end{equation}
Since the spacetime is static $\xi _{i}$ becomes a time-like Killing vector 
and the Lie derivative of the connection present in \eq{Paper1:Eq18}
vanishes. Therefore, in this foliation, holographic equipartition should hold. 
To verify this explicity, we start by calculating 
surface degrees of freedom. From \eq{Paper1:Eq15} the surface degrees of freedom turns out to be:
\begin{equation}\label{Paper1:Eq30}
N_{sur}\equiv A=\int _{\partial \mathcal{R}} \sqrt{\sigma}d^{2}x=4\pi r^{2}
\end{equation}
Again the bulk degree of freedom can be obtained from \eq{Paper1:Eq17} as:
\begin{eqnarray}\label{Paper1:Eq31}
N_{bulk}&=&4\pi \frac{\frac{8\pi}{3}\rho r^{3}}{rl^{-2}}
\end{eqnarray}
Note that the $\epsilon$ factor 
in the definition of the bulk degrees of freedom, keeps it positive, even though the Komar energy density is negative. Then in de Sitter spacetime we have $8\pi \rho =(3/l^{2})$ from which 
we readily observe that:
\begin{equation}\label{Paper1:Eq32}
N_{bulk}=(8\pi \rho)(l^{2}/3)4\pi r^{2} =4\pi r^{2}=N_{sur}
\end{equation}
Hence for de Sitter spacetime in static coordinates  
holographic equipartition does hold as it should. (Alternatively, setting 
$N_{bulk}=N_{sur}$ will lead to the correct identification of $l$ in the metric with source by 
$8\pi \rho =(3/l^{2})$.)

One can easily verify, by explicit computation, how these results generalize to any
static spherically symmetric one, with the line element:
\begin{equation}\label{Paper1:Eq33}
ds^{2}=-f(r)dt^{2}+\frac{dr^{2}}{f(r)}+r^{2}d\Omega ^{2}
\end{equation}
which covers several interesting metrics with horizons.
In this static coordinates the holographic equipartition holds, as can be easily checked. 
A more interesting situation is in the case of geodesic observers in a synchronous frame.
 To check this, we start with a coordinate transformation: 
$\left(t,r,\theta ,\phi \right)\rightarrow \left(\tau ,R,\theta ,\phi \right)$ in which the variables 
are related by the following equations:
\begin{eqnarray}
dt&=&dR -\frac{1}{\sqrt{1-f(r)}}\frac{dr}{f(r)}
\label{Paper1:Eq34a}
\\
dR&=&d\tau +\frac{dr}{\sqrt{1-f(r)}}
\label{Paper1:Eq34b}
\end{eqnarray}
In terms of these newly defined variables the line element reduces 
to the synchronous form:
\begin{equation}\label{Paper1:Eq35}
ds^{2}=-d\tau ^{2}+\left[1-f(r)\right]dR^{2}+r^{2}d\Omega ^{2}
\end{equation}
The comoving observers, having four velocities  $u_{a}=(-1,0,0,0)$ are geodesic observers with zero  acceleration and thus the local Davies-Unruh
temperature also becomes zero. 
We can use \eq{Paper1:Eq21} and \eq{Paper1:Eq23} to describe the evolution. The relevant derivatives are:
\begin{eqnarray}\label{Paper1:Eq36}
\partial _{\tau}g_{RR}&=&-f'(r)\dot{r};~~\partial _{\tau}^{2}g_{RR}=-f'(r)\ddot{r}-f''(r)\dot{r}^{2};
~~\partial _{\tau}g^{RR}=\frac{f'(r)\dot{r}}{\left[1-f(r)\right]^{2}}
\nonumber
\\
\partial _{\tau}g_{\theta \theta}&=&2r\dot{r};~~
\partial _{\tau}^{2}g_{\theta \theta}=2r\ddot{r}+2\dot{r}^{2};~~
\partial _{\tau}g^{\theta \theta}=-\frac{2\dot{r}}{r^{3}}
\nonumber
\\
\partial _{\tau}g_{\phi \phi}&=&2r\dot{r}\sin ^{2}\theta ;~~
\partial _{\tau}^{2}g_{\theta \theta}=\left(2r\ddot{r}+2\dot{r}^{2}\right)\sin ^{2}\theta ;~~
\partial _{\tau}g^{\theta \theta}=-\frac{2\dot{r}}{r^{3}}\frac{1}{\sin ^{2}\theta}
\end{eqnarray}
On substitution of these in \eq{Paper1:Eq23} we obtain the following differential 
equation satisfied by the unknown function $f(r)$:
\begin{equation}\label{Paper1:Eq37}
f''(r)+\frac{2f'(r)}{r}=16\pi \bar{T}_{\tau \tau}=-16 \pi \bar{T}^{0}_{0}
\end{equation}
It can be easily verified that this is the correct field equation in this case 
(see e.g., page 302 of ref. \cite{Padmanabhan2010b}). For example, if we consider the metric of a charged particle with $\bar{T}_{\tau \tau}=Q^{2}/8\pi r^{4}$ above equation can be solved to give $f(r)=1-(2M/r)+(Q^{2}/r^{2})$, which, of course, is the Reissner-Nordstr\"om metric. The description being covariant but foliation dependent, is actually  very 
desirable and inevitable feature from the thermodynamical point of view \cite{Padmanabhan2012c,Padmanabhan2010d}.

\section{Generalization To Lanczos-Lovelock Gravity}\label{Paper1:Sec:LLEqui}

In the previous section we have reviewed, in the context of \EHA, how the departure from holographic equipartition leads to 
the dynamics of the spacetime and have also shown that in static spacetime the surface degrees 
of freedom equals the bulk degrees of freedom. We will now generalize the above description to \LL gravity. 

\subsection{A Brief Introduction to \LL Gravity}
\label{Paper1:Sec:LLEqui:Intro}

Consider, in a D dimensional spacetime, an action functional 
which is made from the metric and the curvature tensor but does not 
contain any derivatives of curvature tensor, such that:
\begin{equation}\label{Paper1:Eq38}
A=\int _{\mathcal{V}}d^{D}x \sqrt{-g}L\left(g^{ab},R^{a}_{~bcd}\right).
\end{equation}
Let us define:
\begin{equation}\label{Paper1:Eq39}
P^{abcd}=\left(\frac{\partial L}{\partial R_{abcd}} \right)_{g_{ij}}
\end{equation}
which has all the algebraic properties of the curvature tensor. We next define another tensor (which is a generalization of Ricci tensor in general relativity) by
\begin{equation}\label{Paper1:Eq40}
\mathcal{R}^{ab}\equiv P^{aijk}R^{b}_{~ijk}.
\end{equation}
This tensor is actually symmetric though the result is nontrivial to prove 
(for this result and more  properties of these tensors see \cite{Padmanabhan2011b}).
The variation of the action functional leads to:
\begin{eqnarray}\label{Paper1:Eq41}
\delta A&=&\delta \int _{\mathcal{V}} d^{D}x \sqrt{-g}L
\nonumber
\\
&=&\int _{\mathcal{V}}d^{D}x \sqrt{-g}E_{ab}\delta g^{ab}
+\int _{\mathcal{V}}d^{D}x \sqrt{-g}\nabla _{j}\delta v^{j}
\end{eqnarray}
where we have the following expression for equation of motion term $E_{ab}$ 
and the boundary term $\delta v^{a}$:
\begin{eqnarray}\label{Paper1:Eq42}
E_{ab}&\equiv & \frac{1}{\sqrt{-g}}\left(\frac{\partial \sqrt{-g}L}{\partial g^{ab}} \right)_{R_{abcd}}
-2\nabla ^{m}\nabla ^{n}P_{amnb}
\nonumber
\\
&=&\mathcal{R}_{ab}-\frac{1}{2}g_{ab}L-2\nabla ^{m}\nabla ^{n}P_{amnb}
\\
\delta v^{j}&=&2P^{ibjd}\nabla _{b}\delta g_{di}-2\delta g_{di}\nabla _{c}P^{ijcd}.
\end{eqnarray}
This is fairly general but we impose the condition that the field equation should be second order in the metric.  
Since the quantity $P^{abcd}$ involves second derivative of the metric, 
the term $\nabla ^{m}\nabla ^{n}P_{amnb}$ in $E_{ab}$ contains fourth order derivative of the 
metric. We can get second order field equation by imposing an extra condition on 
$P^{abcd}$ such that:
\begin{equation}\label{Paper1:Eq43}
\nabla _{a}P^{abcd}=0.
\end{equation}
Thus finding  an action functional which would lead to  equations 
of motion which are second order in the metric reduces to finding scalars 
such that \eq{Paper1:Eq43} is satisfied. Such an action functional is unique and coincides with \LL Lagrangian in 
D dimensions given by \cite{Lanczos1932,Lovelock1971,Padmanabhan2013b,Eddington1924}:
\begin{equation}\label{Paper1:Eq44}
L=\sum _{m}c_{m}L^{(m)}=\sum _{m}c_{m}
\left(\delta ^{aba_{2}b_{2}\ldots a_{m}b_{m}}_{cdc_{2}d_{2}\ldots c_{m}d_{m}}
R^{c_{2}d_{2}}_{a_{2}b_{2}}\ldots R^{c_{m}d_{m}}_{a_{m}b_{m}}\right)R^{cd}_{ab}.
\end{equation}
Due to complete antisymmetry 
in the indices of the determinant tensor, we have in a D dimensional space-time 
the following restriction $2m\leq D$. (Otherwise the determinant tensor would vanish identically.) 
In four dimensions, this property uniquely fixes the result to be the \EHA\ for $m=1$. The nature of \LL models 
at $D=2m$ is of quiet importance as these are known as critical dimensions for a given \LL
term. In these situations the variation of action functional reduces to a pure 
surface term \cite{Padmanabhan2011c}. 

To proceed further, we need the expression for the Noether current in \LL gravity. Recall that the standard result for the  Noether current,  
for diffeomorphism invariance of a Lagrangian $L\left(g^{ab},R^{a}_{~bcd}\right)$, is given by \cite{Padmanabhan2010a}:
\begin{equation}\label{Paper1:Eq45}
16\pi J^{a}=2E^{a}_{b}\xi ^{b}+L\xi ^{a}+\delta _{\xi}v^{a}
\end{equation}
where $E_{ab}$ is defined in \eq{Paper1:Eq42} and $\delta _{\xi}v^{a}$ represents 
the surface term in the Lagrangian variation. 
The following three relations can be used:
\begin{eqnarray}
2E^{a}_{b}\xi ^{b}+L\xi ^{a}&=&2\mathcal{R}^{a}_{b}\xi ^{b}
\label{Paper1:Eq46a}
\\
\delta _{\xi}v^{a}=-\pounds _{\xi}v^{a}
&=&-2\mathcal{R}^{a}_{b}\xi ^{b}+2P^{abdi}\nabla _{b}\nabla_{d}\xi _{i}
\label{Paper1:Eq46b}
\\
\delta _{\xi}v^{i}&=&2P_{a}^{~bci}\pounds _{\xi}\Gamma ^{a}_{bc}
\label{Paper1:Eq46c}
\end{eqnarray}
to express the Noether current  
in two different, useful, forms as follows:
\begin{eqnarray}
16 \pi J^{a}&=&2\mathcal{R}^{a}_{b}\xi ^{b}+\delta _{\xi}v^{a}=2P^{abcd}\nabla _{b}\nabla _{c}\xi _{d}
\label{Paper1:Eq47a}
\\
&=&2\mathcal{R}^{a}_{b}\xi ^{b}+2P_{i}^{~jka}\pounds _{\xi}\Gamma ^{i}_{jk}.
\label{Paper1:Eq47b}
\end{eqnarray}
The corresponding expression for the Noether potential in
\LL gravity is given by \cite{Padmanabhan2010a}:
\begin{eqnarray}\label{Paper1:Eq48}
16 \pi J^{ab}\left(\xi \right)&=&2P^{abcd}\nabla _{c}\xi _{d}.
\end{eqnarray}
We can obtain the entropy of horizons from the relevant Noether charge. In \LL gravity the entropy is defined 
in terms of the tensor $P^{abcd}$ and is known as Wald entropy with the 
expression 
\cite{Wald1993,Padmanabhan2012a,Padmanabhan2013c,Padmanabhan2012b,
Majhi2013,Strominger1996,Ashtekar1998,Bombelli1986}: 
\begin{equation}\label{Paper1:Eq49}
S=-\frac{1}{8}\int d^{D-2}x \sqrt{\sigma}P^{abcd}\mu _{ab}\mu _{cd}\equiv\int d^{D-2}x\ s
\end{equation}
where $\sigma$ is the metric determinant over the $(D-2)$ dimensional hypersurface 
and $\mu _{ab}$ is the bi-normal to the hypersurface. The last equation defines the entropy density $s$ which will be used  frequently 
in our later discussion.
\subsection{Heat Content of Spacetime in \LL Gravity}
\label{Paper1:Sec:LLEqui:Heat}

We will work with the same spacetime foliations defined in \eq{Paper1:Eq01} throughout and thus will use the vectors 
$u^{a}, \xi ^{a}$. 
We begin by performing the same calculation as before, viz. connecting the Noether charge in a volume to the heat content of the boundary. To do this we will start by relating the Noether current for a vector 
$q_{a}$ to that of another vector $f(x)q_{a}=v_{a}$ for any arbitrary 
function $f(x)$. From App. \ref{Paper1:App:NoetherLL} using 
\eq{Paper1:EqB05} we obtain the desired relation as:
\begin{equation}\label{Paper1:Eq50}
16\pi \left\lbrace q_{a}J^{a}(fq)-fq_{a}J^{a}(q)\right\rbrace =\nabla _{b}\left(2P^{abcd}q_{a}q_{d}\nabla _{c}f\right)
\end{equation}
The usefulness of the above equation again originates from the fact that if $q_{a}=\nabla _{a}\phi$ 
then its Noether current vanishes and thus Noether current for $v_{a}=f(x)q_{a}$ acquires 
a particularly simple form. Applying the above result for the two natural vector fields 
$u^{a}$ and $\xi ^{a}$ from \eq{Paper1:EqB13} we obtain the simple relation:
\begin{equation}\label{Paper1:Eq51}
16 \pi u_{a}J^{a}(\xi)=2D_{\alpha}\left(N\chi ^{\alpha}\right)
\end{equation}
where we have introduced a new vector field $\chi ^{a}$ 
given by [see \eq{Paper1:EqB07}]:
\begin{eqnarray}\label{Paper1:Eq52}
\chi ^{a}=-2P^{abcd}u_{b}u_{d}a_{c}
\end{eqnarray}
which satisfies the condition $u_{a}\chi ^{a}=0$ (so that it is a spatial vector) and also has the property: 
$D_{i}\chi ^{i}=\nabla _{i}\chi ^{i}-a_{i}\chi ^{i}$. 
We can integrate \eq{Paper1:Eq51} over $(D-1)$ dimensional 
volume bounded by $N=\textrm{constant}$ surface within 
$t=\textrm{constant}$ hypersurface 
leading to:
\begin{eqnarray}\label{Paper1:Eq53}
\int _{\mathcal{V}}d^{D-1}x \sqrt{h} u^{a}J_{a}\left(\xi \right)
&=&\int _{\partial \mathcal{V}} \frac{d^{D-2}x\sqrt{\sigma}}{8\pi}Nr_{\alpha}\chi ^{\alpha}
\end{eqnarray}
As in \EH, here also the vector $r_{\alpha}$ is the unit normal to the 
$N=\textrm{constant}$ hypersurface. This vector is either parallel or anti-parallel to 
the acceleration four vector such that $r_{\alpha}=\epsilon a_{\alpha}/a$, where $\epsilon =+1$ 
implies parallel to acceleration and vice-versa. With this notion, we obtain the 
following result from the vector field $\chi _{\alpha}$:
\begin{equation}\label{Paper1:Eq54}
\sqrt{\sigma}\frac{Nr_{\alpha}\chi ^{\alpha}}{8\pi}=\epsilon \left(\frac{Na}{2\pi}\right)
\left(\frac{1}{2}\sqrt{\sigma}P^{\alpha b d\beta}r_{\alpha}u_{b}u_{d}r_{\beta}\right)
\end{equation}
The term in brackets is closely related to the entropy density 
of the surface in \LL gravity, defined in \eq{Paper1:Eq49} as \cite{Padmanabhan2010a,Padmanabhan2012a}:
\begin{eqnarray}\label{Paper1:Eq55}
s=-\frac{1}{8}\sqrt{\sigma} P^{abcd}\mu _{ab}\mu _{cd}=\frac{1}{2}\sqrt{\sigma}
P^{\alpha b d\beta}r_{\alpha}u_{b}u_{d}r_{\beta}.
\end{eqnarray}
Using this expression for entropy density  in \eq{Paper1:Eq54} we obtain:
\begin{equation}\label{Paper1:Eq56}
\sqrt{\sigma}\frac{Nr_{\alpha}\chi ^{\alpha}}{8\pi}=\epsilon T_{loc}s
\end{equation}
where $T_{loc}=Na/2\pi$ is the redshifted local Unruh-Davies temperature 
as measured by the observers moving normal to $t=\textrm{constant}$ surface, 
with respect to the local vacuum of freely falling observers. 
We thus see that the results in general relativity has a natural generalization to \LL models.
With all these results, \eq{Paper1:Eq53} reduces to:
\begin{eqnarray}\label{Paper1:Eq57}
\int _{\mathcal{V}}d^{D-1}x \sqrt{h} u^{a}J_{a}\left(\xi \right)
&=&\epsilon \int _{\partial \mathcal{V}}d^{D-2}x~T_{loc}s.
\end{eqnarray}
Thus  \emph{in \LL gravity as well the 
Noether charge in a bulk region is equal to the surface heat content of the boundary}. The similar 
result derived for \EH can be thought of as a special  case of the \LL gravity; the connection between the bulk Noether charge and the surface heat content  
goes way beyond the \EH. This result \textit{is nontrivial because the expression for entropy density in the general \LL models is nontrivial }in contrast with general relativity in which it is just one quarter per unit area.
\subsection{Evolution Equation of Spacetime in \LL Gravity}
\label{Paper1:Sec:LLEqui:Evol}

Let us next generalize our result presented in \eq{start} for \LL models obtaining the dynamical evolution as due to deviation from holographic equipartition.
We will start by substituting the Noether current expression for 
$\xi ^{a}$ as presented in \eq{Paper1:Eq47b} to \eq{Paper1:Eq51} 
which leads to the following result:
\begin{equation}\label{Paper1:Eq58}
2u_{a}P_{i}^{~jka}\pounds _{\xi}\Gamma ^{i}_{jk}=D_{\alpha}\left(2N\chi ^{\alpha}\right)
-2N\mathcal{R}_{ab}u^{a}u^{b}
\end{equation}
Let us first consider the pure \LL\ theory with the 
 $m$th order \LL Lagrangian. (We shall consider the generalization to \LL\ models with a sum of Lagrangians, at the end.) 
Contracting the field equation   $\mathcal{R}_{ab}-(1/2)g_{ab}L=8\pi T_{ab}$ in \LL gravity 
 with $g^{ab}$ we get $L=\left[8\pi\right]/\left[m-(D/2)\right]T$, 
where $D$ is space-time dimension.  Therefore field 
equation can also be rewritten as:
\begin{equation}\label{Paper1:Eq59}
\mathcal{R}_{ab}=8\pi \bar{T}_{ab}
=8\pi \left(T_{ab}-\frac{1}{2}\frac{1}{(D/2)-m}g_{ab}T\right)\equiv 8\pi\bar{T}_{ab}
\end{equation}
Using this and integrating \eq{Paper1:Eq58}
over $(D-1)$ dimensional volume we arrive at:
\begin{equation}\label{Paper1:Eq60}
\int _{\mathcal{R}}\frac{d^{D-1}x\sqrt{h}}{8\pi}2u_{a}P_{i}^{~jka}\pounds _{\xi}\Gamma ^{i}_{jk}
=\int _{\partial \mathcal{R}}\frac{d^{D-2}x\sqrt{\sigma}}{4\pi}N\chi ^{\alpha}r_{\alpha}
-\int _{\mathcal{R}}d^{D-1}x\sqrt{h}2N\bar{T}_{ab}u^{a}u^{b}.
\end{equation}
As before, the $r_{\alpha}$ is the normal to $N=\textrm{constant}$ surface 
within $t=\textrm{constant}$ surface and is either parallel or 
anti-parallel to the four acceleration. The energy momentum term 
can be written in an identical fashion by using the Komar energy density, 
defined as: $\rho _{Komar}=2N\bar{T}_{ab}u^{a}u^{b}$. We can proceed using \eq{Paper1:Eq54}, 
which on substitution into \eq{Paper1:Eq60} leads to:
\begin{equation}\label{Paper1:Eq61}
\int _{\mathcal{R}}\frac{d^{D-1}x\sqrt{h}}{8\pi}2u_{a}P_{i}^{~jka}\pounds _{\xi}\Gamma ^{i}_{jk}
=-2\epsilon\int _{\partial \mathcal{R}}d^{D-2}x\sqrt{\sigma}
P^{\alpha b \beta d}r_{\alpha}u_{b}r_{\beta}u_{d}\left(\frac{1}{2}T_{loc}\right)
-\int _{\mathcal{R}}d^{D-1}x\sqrt{h}\rho _{Komar}
\end{equation}
Rest of the analysis requires proper definition of $N_{sur},N_{bulk}$ etc which we do in analogy with the case of general relativity.
The number of surface degrees of freedom is defined as four times the entropy as in the case of general relativity:
\begin{equation}\label{Paper1:Eq62}
N_{sur}\equiv 4S=2\int _{\partial \mathcal{R}}d^{D-2}x\sqrt{\sigma}
P^{\alpha b d\beta }r_{\alpha}u_{b}u_{d}r_{\beta}
\end{equation}
The average temperature is properly defined using the surface degrees of freedom as the local weights leading to ensure that the total heat content is reproduced:
\begin{equation}
\frac{1}{2}N_{sur} k_BT_{avg}=\frac{1}{2}\int dN_{sur} k_BT_{loc};\qquad
 T_{avg}S=\int T_{loc} dS.
\end{equation} 
This result can be written, more explicitly as:
\begin{equation}\label{Paper1:Eq63}
T_{avg}=\frac{\int _{\partial \mathcal{R}}d^{D-2}x\sqrt{\sigma}
P^{\alpha b \beta d}r_{\alpha}u_{b}r_{\beta}u_{d}T_{loc}}
{\int _{\partial \mathcal{R}}d^{D-2}x\sqrt{\sigma}
P^{\alpha b \beta d}r_{\alpha}u_{b}r_{\beta}u_{d}}=\frac{1}{S}\int dS T_{loc}
=\frac{1}{N_{sur}}\int dN_{sur} T_{loc}
\end{equation}
Once $T_{avg}$ is defined, the number of bulk degrees of freedom is given by the equipartition value:
\begin{equation}\label{Paper1:Eq64}
N_{bulk}=\frac{\epsilon}{(1/2)T_{avg}}\int _{\mathcal{R}}d^{D-1}x\sqrt{h}\rho _{Komar}
\end{equation}
with $\epsilon$ included (as in general relativity), to ensure that $N_{bulk}$ is always positive. 
Inserting \eqs{Paper1:Eq62}, (\ref{Paper1:Eq63}) and (\ref{Paper1:Eq64}) in 
 \eq{Paper1:Eq61}
we find that the dynamical evolution of the spacetime in \LL gravity is
determined by the following relation:
\begin{equation}\label{Paper1:Eq65}
\int _{\mathcal{R}}\frac{d^{D-1}x\sqrt{h}}{8\pi}2u_{a}P_{i}^{~jka}\pounds _{\xi}\Gamma ^{i}_{jk}
=\epsilon \left(\frac{1}{2}T_{avg}\right)\left(N_{sur}-N_{bulk}\right)
\end{equation}
which is direct generalization of the corresponding result for general relativity.

For a static spacetime  the Lie variation of connection vanishes 
as $\xi ^{a}$ becomes a time-like Killing vector. Hence in that situation we have, even in in \LL gravity, the holographic equipartition given by:
\begin{equation}\label{Paper1:Eq66}
N_{sur}=N_{bulk}
\end{equation} 
(This result has been obtained earlier in terms of equipartition energies in ref. \cite{Padmanabhan2010c}.) 
When the foliation leads to time dependent metric, the departure from holographic equipartition 
drives dynamical evolution of the metric through  the Lie derivative term on the 
left hand side of \eq{Paper1:Eq65}. 

The above result was derived for $m$th order \LL Lagrangian. The definition of $\bar{T}_{ab}, \rho_{Komar}$ and $N_{bulk}$ introduces the $m$ dependence though the expression for $\mathcal{R}_{ab}$ in \eq{Paper1:Eq59}. If, instead, we consider a \LL\ Lagrangian made of a sum of Langrangians with different $m$, then the equation of motion, 
$\mathcal{R}_{ab}-(1/2)g_{ab}L=8\pi T_{ab}$ on contraction with $g^{ab}$ leads to the result: 
\begin{equation}
\sum _{m}c_{m}\left[m-(D/2)\right]L_{(m)}=8\pi T                                                \end{equation}  
which cannot be solved in closed form for $L$ in terms of $T$. However, one can take care of this issue by redefining $\rho_{Komar}$ and $N_{bulk}$ formally in terms of $\mathcal{R}_{ab}$. That is, we define the Komar energy density as: 
$\rho =2N(\mathcal{R}_{ab}/8\pi)u^{a}u^{b}$ and then the bulk degrees of freedom 
reduces to the following form:
\begin{equation}\label{Paper1:Eq69}
N_{bulk}=\frac{\epsilon}{(1/2)T_{avg}}\int _{\mathcal{R}}d^{D-1}x\sqrt{h}\rho
\end{equation}
Then we again obtain the same result:
\begin{equation}\label{Paper1:Eq70}
\int _{\mathcal{R}}\frac{d^{D-1}x\sqrt{h}}{8\pi}2u_{a}P_{i}^{~jka}\pounds _{\xi}\Gamma ^{i}_{jk}
=\epsilon \left(\frac{1}{2}T_{avg}\right)\left(N_{sur}-N_{bulk}\right)
\end{equation}
with the understanding that, for a given model, one should re-express the variables in terms of $T_{ab}$.

The above results provide a direct connection between evolution of spacetime and 
departure from holographic equipartition. The results also encode the holographic behavior 
of gravity by introducing naturally defined bulk and surface degrees of freedom. 
The difference between the description of evolution along these lines and that of standard field equation 
$\mathcal{R}_{ab}-(1/2)g_{ab}L=8\pi T_{ab}$ is  the following: For the standard 
gravitational field equations the left hand side \emph{does not} have a clear physical meaning. There is also 
 no distinction between static and dynamic spacetime and hence the standard treatment  cannot answer 
the question: what drives the time-dependence of the metric? The answer  is obviously not 
$T_{ab}$ since we can obtain time dependent solutions even when $T_{ab}=0$ and static solutions with $T_{ab}\neq0$.
In contrast the evolution depicted in \eq{Paper1:Eq70} addresses all these issues and we have a natural separation between static and evolving metrics via 
holographic equipartition. 
When the surface and bulk degrees of freedom are unequal, resulting in 
departure from holographic equipartition, it drives the time-dependence of the metric. Thus the driving 
force behind dynamical evolution of spacetime is the departure from holographic equipartition, 
providing a physically transparent statement about spacetime dynamics.

\section{Discussion}\label{Paper1:Sec:Discussion}

Our aim in this work was to consider the relationship between the Noether 
current and gravitational dynamics in a useful manner. Noether currents 
can be thought of as originating from mathematical identities in differential 
geometry, with \emph{no connection to the diffeomorphism invariance 
of gravitational action} \cite{Padmanabhan2013a}. This result holds not only in \EH  but also in \LL gravity (see Appendix \ref{Paper1:App:DiffNoether} ). 

Even though such conserved currents can be associated with 
any vector field, the time development vectors are always special. This 
is the motivation for introducing the vector $\xi ^{a}$  
in the  spacetime through \eq{Paper1:Eq01}. 
The vector $\xi ^{a}$ is parallel to velocity vector $u^{a}$ for fundamental 
observers and represents proper time flow normal to $t=\textrm{constant}$ surface. 
As we saw, its Noether charge and current associated with this vector have elegant and physically interesting thermodynamic interpretation. We showed  that, for the vector field $\xi ^{a}$ 
in \LL gravity in arbitrary spacetime dimension, \emph{total Noether charge} in any bulk 
volume $\mathcal{V}$, bounded by constant lapse surface, equals \emph{the heat content 
of the boundary surface}. Also the equipartition energy of the surface equals twice the Noether 
charge. While defining the heat content, we have 
used local Unruh-Davies temperature and Wald entropy. This result holds for \LL gravity 
of all orders and does not rely on static spacetime or existence of Killing vector like criteria. 

The above identification allow us to study holographic equipartition for static spacetime 
and relate the time evolution of the metric as  due to departure 
from holographic equipartition. With a suitable and natural definition for the degrees of freedom in the surface 
and in the bulk, we find that for static spacetimes (described in the natural foliation) the surface and the bulk degrees of freedom are equal in number
yielding holographic equipartition. It is the departure from this holographic equipartition that 
drives spacetime evolution. This result holds not only in \EH  but also 
in \LL gravity. 

All the results derived above are generally covariant but they do depend on the
foliation. This implies that these results depend on observers and their acceleration which is inevitable since the Davies-Unruh temperature is intrinsically observer dependent. 
Since the dynamical evolution is connected to 
thermodynamic concepts in this approach, different observers \textit{must} perceive the dynamical evolution 
differently. For example, the de Sitter spacetime is time dependent when written in 
synchronous frame, becomes time independent in static spherically symmetric 
coordinate. Our description adapts naturally to the two different situations.
\section*{Acknowledgement}

Research of T.P is partially supported by J.C. Bose research grant of DST, Govt. of India. 
Research of S.C 
is funded by SPM Fellowship from CSIR, Govt. of India. 
S.C also likes to thank Krishnamohan Parattu, Suprit Singh and Kinjalk Lochan 
for helpful discussions. We thank Naresh Dadhich for useful comments.
\appendix

\section{Calculational Details}\label{Paper1:App}

Some calculations are not presented in an explicit format in the main text, which 
would affect the flow of ideas in the paper. Most of these relations exist 
in the literature; however we collect the derivations together here with the hope 
that they will be useful to the reader.

\subsection{Derivation of Noether Current from differential Identities in \LL Gravity}
\label{Paper1:App:DiffNoether}

In this section  the Noether current for \LL\ gravity will be derived starting from 
identities in differential geometry without using any difeomorphism invariance of action principles.
The conceptual importance of this approach has already been emphasized in ref. \cite{Padmanabhan2013a}, in 
the context of Einstein gravity, and we shall generalize the result for \LL\ models. 
We start with the fact that the covariant derivative of any vector field can be decomposed 
into a symmetric and an antisymmetric part. From 
the antisymmetric part we can define another antisymmetric tensor field as,
\begin{equation}\label{Paper1:EqC01}
16 \pi J^{aj}=2P^{ajki}\nabla _{k}v_{i}=P^{ajki}\left(\nabla _{k}v_{i}-\nabla _{i}v_{k}\right)
\end{equation}
It is evident from the antisymmetry of $P^{abcd}$ that a conserved current exists such 
that, $J^{a}=\nabla _{j}J^{aj}$. We recall the identities: 
\begin{equation}\label{Paper1:EqC02}
\left(\nabla _{j}\nabla _{k}-\nabla _{k}\nabla _{j} \right)v^{i}=R^{i}_{~cjk}v^{c}
\end{equation}
and,
\begin{equation}\label{Paper1:EqC03}
\mathcal{L}_{v}\Gamma ^{i}_{jk}=\nabla _{j}\nabla _{k}v^{i}-R^{i}_{~kjm}v^{m}
\end{equation}
and use them in the definition in \eq{Paper1:Eq40} to get:
\begin{eqnarray}\label{Paper1:EqC04}
\mathcal{R}^{ab}v_{b}&=&P^{aijk}R^{b}_{~ijk}v_{b}
=-P^{aijk}\left( \nabla _{j}\nabla _{k}-\nabla _{k}\nabla _{j}\right)v_{i}
\nonumber
\\
&=&P^{aijk}\nabla _{k}\nabla _{j}v_{i}+\left(P^{akij}+P^{ajki}\right)\nabla _{j}\nabla _{k}v_{i}
\nonumber
\\
&=&P^{aijk}\nabla _{k}\nabla _{j}v_{i}+P^{akij}\nabla _{j}\nabla _{k}v_{i}
+\nabla _{j}\left(P^{ajki}\nabla _{k}v_{i} \right)
\end{eqnarray}
where in the second line we have used the identity, $P^{a(bcd)}=0$. 
Then from \eq{Paper1:EqC01} we 
 obtain:
\begin{eqnarray}\label{Paper1:EqC05}
16 \pi J^{a}&=&2\mathcal{R}^{ab}v_{b}-2P^{aijk}\nabla _{k}\nabla _{j}v_{i}-2P^{akij}\nabla _{j}\nabla _{k}v_{i}
\nonumber
\\
&=&2\mathcal{R}^{ab}v_{b}+2P_{i}^{~ajk}\nabla _{k}\nabla _{j}v^{i}-2P_{i}^{~jak}\nabla _{j}\nabla _{k}v^{i}
\nonumber
\\
&=&2\mathcal{R}^{ab}v_{b}+2P_{i}^{~ajk}\left(\mathcal{L}_{v}\Gamma ^{i}_{kj}+R^{i}_{~jkm}v^{m} \right)
-2P_{i}^{~jak}\left(\mathcal{L}_{v}\Gamma ^{i}_{jk}+R^{i}_{~kjm}v^{m}\right)
\nonumber
\\
&=&2\mathcal{R}^{ab}v_{b}+2P_{i}^{~jka}\mathcal{L}_{v}\Gamma ^{i}_{jk}
\end{eqnarray}
while arriving at the third line we have used \eq{Paper1:EqC03} and for the last line we have used 
the fact that, $P^{ijak}R_{ikjm}=P^{akij}R_{ikjm}=-P^{kaij}R_{ikjm}=P^{kaij}R_{kijm}$. 
Thus \eq{Paper1:Eq47b} can be derived without any reference to the diffeomorphism invariance of the gravitational action, using only the identities in 
differential geometry and various symmetry properties. 
\subsection{Identities Regarding Noether current in \LL Action}
\label{Paper1:App:NoetherLL}

The Noether potential $J^{ab}$ is antisymmetric in $(a,b)$ and from its expression 
given by \eq{Paper1:Eq04} it 
is evident that $J^{ab}(q)$ would identically vanish for $q_{a}=\nabla _{a}\phi$. We will use 
the above fact in order to obtain a relation between the Noether current for two vector fields $q_{a}$ and $v_{a}$ 
connected by $v_{a}=f(x)q_{a}$. This result, in the case of general relativity 
is detailed in ref. \cite{Padmanabhan2013a}.
Expanding the expression for Noether current for $v_{a}=fq_{a}$ 
and taking dot product with $q_{a}$ along with subtracting the Noether current for $q_{a}$ 
one can show that:
\begin{eqnarray}\label{Paper1:EqA02}
16 \pi \left\lbrace q_{a}J^{a}(fq)-fq_{a}J^{a}(q)\right\rbrace
=\nabla _{b}\left[\left(q^{a}q^{b}-q^{2}g^{ab}\right)\nabla _{a}f\right].
\end{eqnarray}
This is the result  used  in the main text. 
Using this result it is easy to detemine the Noether currents for $u_{a}=-N\nabla _{a}t$ and 
 $\xi _{a}=Nu_{a}$. 
 Using \eq{Paper1:EqA02} with $q_{a}=-u_{a}/N$ and 
$f=-N$ we obtain:
\begin{equation}\label{Paper1:EqA06}
16 \pi u_{a}J^{a}(u)=\nabla _{i}a^{i}-a^{2}=D_{\alpha}a^{\alpha}
\end{equation}
where the acceleration is defined as:
\begin{equation}\label{Paper1:EqA04}
a_{j}=u^{i}\nabla _{i}u_{j}
=\left(u^{i}\nabla _{i}N\right)\frac{u_{j}}{N}+Nu^{i}\nabla _{j}\left(\frac{u_{i}}{N}\right)
=h^{i}_{j}\frac{\nabla _{i}N}{N}.
\end{equation}
Next in order to obtain the Noether current for $\xi ^{a}$ we use \eq{Paper1:EqA02} with 
$q_{a}=u_{a}$ and $f=N$ leading to:
\begin{equation}\label{Paper1:EqA08}
16 \pi u_{a}J^{a}(\xi)=Nu_{a}J^{a}(u)+\nabla _{j}\left(Na^{j}\right)=2N\nabla _{j}a^{j}=
D_{\alpha}\left(2Na^{\alpha}\right)
\end{equation}
which is the desired relation 
in \eq{Paper1:Eq06}.

In \EH the quantity $u_{a}g^{ij}\pounds _{\xi}N^{a}_{ij}$ can be evaluated 
in terms of the extrinsic curvature \cite{Padmanabhan2010b}. 
Then from the standard identity 
\begin{eqnarray}\label{Paper1:EqA11}
\nabla _{i}a^{i}-R_{ab}u^{a}u^{b}=K_{ij}K^{ij}-u^{a}\nabla _{a}K
\end{eqnarray}
we obtain:
\begin{equation}\label{Paper1:EqA12}
u_{a}g^{ij}\pounds _{\xi}N^{a}_{ij}=2N\left(\nabla _{i}a^{i}-R_{ab}u^{a}u^{b}\right)
=2N\left(K_{ij}K^{ij}-u^{a}\nabla _{a}K\right)
\end{equation}

Next we will generalize the above results to \LL gravity. 
For that purpose we note that even in \LL gravity the Noether potential $J^{ab}$ for a vector field 
$q_{a}=\nabla _{a}f$ vanishes identically. Thus the Noether current for a vector 
field $v_{a}=f(x)q_{a}$ can be decomposed as:
\begin{eqnarray}\label{Paper1:EqB01}
16 \pi J^{ab}(v)&=&2P^{abcd}\nabla _{c}\left(fq_{d}\right)
\nonumber
\\
&=&2P^{abcd}q_{d}\nabla _{c}f+2fP^{abcd}\nabla _{c}q_{d}
\end{eqnarray}
Then the corresponding Noether current has the following expression:
\begin{eqnarray}\label{Paper1:EqB02}
16 \pi J^{a}(v)&=&2P^{abcd}\nabla _{b}\left(q_{d}\nabla _{c}f\right)
+2P^{abcd}\nabla _{b}\left(f\nabla _{c}q_{d}\right)
\nonumber
\\
&=&2P^{abcd}q_{d}\nabla _{b}\nabla _{c}f+2P^{abcd}\nabla _{c}f\nabla _{b}q_{d}
+2P^{abcd}\nabla _{b}f\nabla _{c}q_{d}+2fP^{abcd}\nabla _{b}\nabla _{c}q_{d}
\end{eqnarray}
From the above equation we readily arrive at:
\begin{eqnarray}\label{Paper1:EqB03}
16 \pi \left\lbrace J^{a}(v)-fJ^{a}(q)\right\rbrace &=&
2P^{abcd}q_{d}\nabla _{b}\nabla _{c}f+2P^{abcd}\nabla _{c}f\nabla _{b}q_{d}
+2P^{abcd}\nabla _{b}f\nabla _{c}q_{d}
\nonumber
\\
&=&P^{abcd}\nabla _{b}A_{cd}+16 \pi J^{ab}(q)\nabla _{b}f
\end{eqnarray}
where we have defined the antisymmetric tensor $A_{cd}$ as $A_{cd}=q_{d}\nabla _{c}f-q_{c}\nabla _{d}f$. 
Now consider the following result: $q_{a}\nabla _{b}A_{cd}=\nabla _{b}\left(q_{a}A_{cd}\right)
-A_{cd}\nabla _{b}q_{a}$ which leads to:
\begin{eqnarray}\label{Paper1:EqB04}
P^{abcd}q_{a}\nabla _{b}A_{cd}&=&\nabla _{b}\left(P^{abcd}q_{a}A_{cd}\right)
-2P^{abcd}q_{d}\nabla _{c}f\nabla _{b}q_{a}
\nonumber
\\
&=&\nabla _{b}\left(P^{abcd}q_{a}A_{cd}\right)-16 \pi q_{a}J^{ab}\left(q\right)\nabla _{b}f
\end{eqnarray}
Then \eq{Paper1:EqB03} can be rewritten in the following manner:
\begin{eqnarray}\label{Paper1:EqB05}
16 \pi \left\lbrace q_{a}J^{a}(fq)-fq_{a}J^{a}(q)\right \rbrace &=&
16 \pi J^{ab}(q)\nabla _{b}fq_{a}
+\nabla _{b}\left(P^{abcd}q_{a}A_{cd}\right)-16 \pi q_{a}J^{ab}(q)\nabla _{b}f
\nonumber
\\
&=&\nabla _{b}\left(2P^{abcd}q_{a}q_{d}\nabla _{c}f\right)
\end{eqnarray}
It can be easily verified that in the \EH limit 
$P^{abcd}=Q^{abcd}=(1/2)\left(g^{ac}g^{bd}-g^{ad}g^{bc}\right)$, 
under which the above equation reduces to 
\eq{Paper1:EqA02}. 

Applying the above equation to $u_{a}=-N\nabla _{a}t$ 
with $q_{a}=\nabla _{a}t=-u_{a}/N$ and $f=-N$
we arrive at:
\begin{equation}\label{Paper1:EqB06}
16 \pi u_{a}J^{a}\left(u\right)=2N\nabla _{b}\left(P^{abcd}u_{a}u_{d}\frac{\nabla _{c}N}{N^{2}}\right)
\end{equation}
In order to proceed we define a new vector field such that:
\begin{eqnarray}\label{Paper1:EqB07}
\chi ^{a}&=&-2P^{abcd}u_{b}u_{d}\frac{\nabla _{c}N}{N}
\nonumber
\\
&=&-2P^{abcd}u_{b}u_{d}\left(a_{c}-\frac{1}{N}u_{c}u^{j}\nabla _{j}N\right)
\nonumber
\\
&=&-2P^{abcd}u_{b}a_{c}u_{d}
\end{eqnarray}
Note that in the \EH limit this vector reduces to the acceleration four vector as 
follows:
\begin{equation}\label{Paper1:EqB08}
\chi ^{a}=-2P^{abcd}u_{b}a_{c}u_{d}=-\left(g^{ac}g^{bd}-g^{ad}g^{bc}\right)u_{b}a_{c}u_{d}
=-u^{b}u_{b}a^{a}+u^{b}a_{b}u^{a}=a^{a}
\end{equation}
Also just as in the case of acceleration for the vector $\chi ^{a}$ as well we have:
\begin{eqnarray}\label{Paper1:EqB09}
u_{a}\chi ^{a}&=&-2aP^{ab\beta d}u_{a}u_{b}r_{\beta}u_{d}=0
\end{eqnarray} 
where antisymmetry of $P^{abcd}$ in the first two components has been used. 
We can also have the following relation for the vector field $\chi ^{a}$:
\begin{equation}\label{Paper1:EqB10}
Na_{b}\chi ^{b}=\chi ^{b}\nabla _{b}N +\chi ^{b}u_{b}u^{j}\nabla _{j}N=\chi ^{b}\nabla _{b}N
\end{equation}
where we have used the relation $u_{a}\chi ^{a}=0$ from \eq{Paper1:EqB09}. 
Thus \eq{Paper1:EqB06} can be written in terms of the 
newly defined vector field $\chi ^{a}$ in the following way:
\begin{eqnarray}\label{Paper1:EqB11}
16 \pi u_{a}J^{a}\left(u\right)&=&N\nabla _{b}\left(\frac{\chi ^{b}}{N}\right)
\nonumber
\\
&=&\nabla _{b}\chi ^{b}-\frac{\nabla _{b}N}{N}\chi ^{b}
\nonumber
\\
&=&D_{\alpha}\chi ^{\alpha}
\end{eqnarray}
The last relation follows from the fact that:
\begin{equation}\label{Paper1:EqB12}
D_{\alpha}\chi ^{\alpha}=
D_{b}\chi ^{b}=\nabla _{b}\chi ^{b}-a_{b}\chi ^{b}=\nabla _{b}\chi ^{b}-\frac{\nabla _{b}N}{N}\chi ^{b}
\end{equation}
Then it is straightforward to get the Noether current for $\xi ^{a}$ by 
using $q_{a}=u_{a}$ and $f=N$ in \eq{Paper1:EqB05} with \eq{Paper1:EqB11} as:
\begin{eqnarray}\label{Paper1:EqB13}
16 \pi u^{a}J_{a}\left(\xi \right)&=&16 \pi Nu_{a}J^{a}(u)+\nabla_{b}\left(N\chi ^{b}\right)
\nonumber
\\
&=&ND_{\alpha}\chi ^{\alpha}+\nabla _{b}\left(N\chi ^{b}\right)
\nonumber
\\
&=&D_{\alpha}\left(2N\chi ^{\alpha}\right)
\end{eqnarray}
Here also we have used the following identity:
\begin{eqnarray}
D_{\alpha}\left(N\chi ^{\alpha}\right)&=&\left(g^{ij}+u^{i}u^{j}\right)\nabla _{i}\left(N\chi _{j}\right)
\nonumber
\\
&=&\nabla _{i}\left(N\chi ^{i}\right)+u^{i}u^{j}\nabla _{i}\left(N\chi _{j}\right)
\nonumber
\\
&=&N\nabla _{i}\chi ^{i}+N\chi ^{i}a_{i}-N\chi ^{j}\left(u^{i}\nabla _{i}u_{j}\right)
\nonumber
\\
&=&N\nabla _{i}\chi ^{i}
\end{eqnarray}
Thus we have derived the desired relation for the Noether current 
of the vector field $\xi _{a}$ and it turns out to have identical structure 
as that of \EH action with $\chi ^{a}$ playing the role of four acceleration.



\end{document}